\documentstyle[prl,aps,psfig]{revtex}

\begin{document}
\draft

\twocolumn[
\hsize\textwidth\columnwidth\hsize\csname @twocolumnfalse\endcsname

\title{Distribution of time-constants for tunneling  through a $1D$ Disordered Chain}
 
\author{C.J. Bolton-Heaton$^*$, C.J. Lambert$^*$, Vladimir I.  Falko$^*$, V. Prigodin$^{\dagger}$ and A.J. Epstein ${\dagger}$}
 
\address{$^*$ Department of Physics,
Lancaster University, Lancaster LA1 4YB, UK\\
$^{\dagger}$ Physics Department, The Ohio State University, Columbus, OH 43210-1106, USA }
 \date{\today}
  
\maketitle 

\begin{abstract} The dynamics of electronic tunneling through a
 disordered $1D$ chain of finite length is considered. We
  calculate  distributions of the transmission coefficient $T$,
Wigner delay time and, $\tau_\phi$ and the transport time,
$\tau_t=T\tau_\phi$. The central bodies of 
 these distributions have a power-law form, what can be
understood in terms of the resonant tunneling through 
localised states. 
\end{abstract}
]

During the past decade, transport properties of phase-coherent disordered
low-dimensional conductors attracted much interest. The transmission
properties of complex structures with a large number of internal degrees of
freedom have been intensively studied using random matrix theory \cite
{Altshuler96}, and a universal distribution for transmission coefficients
through metallic systems has been derived \cite{Jalabert92}. The
one-dimensional localization problem is exactly solvable, and a complete
description of the distribution of transmission coefficients, $T$ and
localization properties of single-electron wave functions in disordered 1D
and quasi-1D wires is now available \cite{Fisher80,BeenakkerReview}. In
particular, it has been found that the inverse localization radius, $\alpha $
of single-particle localized states by disorder in a 1D chain has a normal
distribution with width inversely proportional to the chain length, $L$: 
\begin{equation}
P(\alpha )\sim \exp \left\{ -L\frac{(\alpha -\alpha _{0})^{2}}{2\alpha _{0}}%
\right\} .  \label{p-alpha}
\end{equation}
Here, $\alpha _{0}^{-1}$ is the most probable localization radius which for
a weakly disordered system is given by $\alpha _{0}^{-1}=4l$, where $l$ is
the mean free path. This is equivalent to a log-normal distribution of the
transmission coefficients, $T\sim \exp (-\alpha L)$, 
\begin{equation}
P(T)\sim \frac{1}{T}\exp \left\{ -\frac{\left( \ln (1/T)-2L\alpha
_{0}\right) ^{2}}{8L\alpha _{0}}\right\} .  \label{p-T}
\end{equation}

Recently, random matrix theory was applied to the problem of the dynamical
electric response of mesoscopic conductors and disordered wires in the
localized regime. Here the problem involves understanding the distribution
of the Wigner delay time, $\tau _{\phi }$ which carries information about
the life-time of carriers in the resonant states responsible for
transmission through a weakly couple quantum dot, or through a disordered
wire in the localized regime. The Wigner delay time is related to the\
energy dependence of the phase shift $\theta (\epsilon )$ of a wave passing
through a disordered wire or a quantum resonator, and is given by 
\begin{equation}
\tau _{\phi }=\frac{d\theta (\epsilon )}{d\epsilon }=\frac{1}{2i}\frac{d}{%
d\epsilon }\ln \frac{G_{R}(\epsilon ,L)}{G_{A}(\epsilon ,L)}{.}
\label{wignertime}
\end{equation}
This equation also expresses the delay time in terms of retarded and
advanced Green functions, $G_{R,A}(\epsilon ,L)=\sum_{\nu }\psi _{\nu
}^{\ast }(0)\psi _{\nu }(L)/\left[ \epsilon -\epsilon _{\nu }\pm i0\right] $%
, which links $\tau _{\phi }$ to the density of states of a system, $\tau
_{\phi }=\pi \int_{0}^{L}dx\nu (\epsilon ,x)$. The Wigner delay time in
zero-dimensional mesoscopic systems modelled using random matrix theory has
been studied \cite{Fyodorov96} within the zero-dimensional $\sigma $-model
approach, and has been shown to have a universal
distribution. In the present paper, we analyze a related dynamical
characteristic of a disordered conductor, namely, the transport time $\tau
_{t}$ defined as the delay time weighted by the transmission coefficient, 
\begin{equation}
\tau _{t}=T\tau _{\phi }.  \label{transporttime}
\end{equation}
This quantity characterizes the ability of a resonant state to provide a
dynamical response to an external ac-electric field. Using the
Landauer-Buttiker approach, the imaginary part of the dimensionless
ac-conductance $g=G(\omega )/\frac{2e^{2}}{h}$ of a single-channel
mesoscopic wire \cite{Pastawski92,Buttiker93}, 
\[
g=T(\omega )=v_{F}^{2}G_{R}(\epsilon +\omega /2,L)G_{A}(\epsilon -\omega
/2,L), 
\]
can be represented as 
\[
{\rm Im}g(\omega )=T{\rm Im}
[1-i\omega \tau _{\phi }+...]=-\omega \tau _{t}, 
\]
so that the transport time, $\tau _{t}$ is a directly measurable quantity,
which can be also interpreted as the dielectric response function of an
almost insulating 1D wire, when ${\rm Re}G\rightarrow 0$. 
In the present paper, we report the results of numerical
studies and a qualitative asymptotic analysis of the transport-time
distribution function $P(\tau _{t})$ in the localized regime of 1D
disordered wires. To anticipate a little, we find that the distribution of
this quantity is affected by correlations between the value of the Wigner
delay time and the transmission coefficient of resonances via localized
states. Using information about the distribution of the localization radia
in Eq. (\ref{p-alpha}) and about the energetic widths of individual
resonances, we show that the central body of the distribution of $\tau_t$, 
which corresponds to $-2/3<(1/z)\ln (\tau _{t}/\tau )<1/3 $, where $%
z=L\alpha _{0}$, is given by the power-law asymptotic, 
\begin{equation}
P(\tau _{t})\sim \tau ^{-1}e^{-z/2}\left( \frac{\tau }{\tau _{t}}\right)
^{4/3},  \label{p-tauT1}
\end{equation}
in complete agreement with our numerical simulations. The tail of short
times $\tau _{t}$, $(1/z)\ln (\tau _{t}/\tau )<-2/3$ decays in the
logarithmically normal way, 
\begin{equation}
P(\tau _{t})\sim \tau ^{-1}e^{-z/2}\left( \frac{\tau _{t}}{\tau }\right)
^{-(3/2)+(1/(8z))\ln (\tau /\tau _{t})},  \label{p-tauT2}
\end{equation}
whereas for $(1/z)\ln (\tau _{t}/\tau )>1/3$, 
\begin{equation}
P(\tau _{t})\sim \tau ^{-1}e^{-z/2}\left( \frac{\tau }{\tau _{t}}\right)
^{1+(1/(2z))\ln (\tau _{t}/\tau )}.\;  \label{p-tauT3}
\end{equation}

Below, we show how Eqs. (\ref{p-tauWigner}-\ref{p-tauT3}) can be obtained,
And describe the numerical procedure used to determine the distribution of $%
\tau _{\phi }$ and $\tau _{t}$. We begin with an analysis of the Wigner
delay time and introduce the numerically-studied model. In agreement with
the result of Ref. \cite{WignTimeDisrt}, we show, both analytically and
numerically, that the body of the distribution function $P(\tau _{\phi })$
is dominated by the inverse-square-law asymptotic 
\begin{equation}
P(\tau _{\phi })\sim \tau /\tau _{\phi }^{2},\;{\rm at}\;\tau _{\phi }>\tau ,
\label{p-tauWigner}
\end{equation}
where $\tau $ is the mean free path time. These results provide a check both
of both the analytical method and the numerics and are followed by an the
analysis of the transport-time distribution, which is the central goal of
this paper.

To obtain the distribution of $\tau_\phi$, we note that in the absence of
disorder in a 1D wire, $T=1$ and $\tau _{\phi }=\tau _{t}=L/v_{F}$, $v_{F}$
being the electron Fermi velocity which determines the ballistic time of
flight of an electron through the chain. For a weakly disordered chain
characterized by a mean free path $l=v\tau $ or scattering time $\tau \ll
1/\epsilon $, the transmission $T$, $\tau _{\phi }$ and $\tau _{t}$ are
random variables. In a long wire $L\gg l$, where localization is strong,
transmission can be viewed as being the result of tunneling through resonant
levels, each characterized by its energy, $\epsilon _{0}$ and decay width, $%
\gamma =\gamma _{1}+\gamma _{2}$ determined by the electron escape rates $%
\gamma _{1,2}$ into the left and right contacts. In the exponential
localization regime $L\gg l$, the tunnelling rates associated with a
resonant state peaked at $x<L$ (calculated from the left end of the chain)
are of order 
\begin{equation}
\gamma _{1}=\tau ^{-1}e^{-2\alpha _{1}x}~~~~{\rm and}~~~~\gamma _{2}=\tau
^{-1}e^{-2\alpha _{2}(L-x)},  \label{gamma}
\end{equation}
where $\alpha _{1}$ and $\alpha _{2}$ are two independently fluctuating
inverse localization radia of the wave function tail on the left and right
hand sides of the resonance. Note that since the resonance width falls
exponentially with the wire length, whereas the mean level spacing $\Delta
=1/\nu L$ is only inverse proportional to $L$, the assumption $L\gg l$
allows us to distinguish between resonances and to consider each resonant
state as a slightly broadened discrete level.

Therefore, at each value of the energy, the ac-transmission through the
disordered 1D chain can be described using the Bright-Wigner formula, 
\[
T(\epsilon )=\frac{4\gamma _{1}\gamma _{2}}{(\epsilon -\epsilon
_{0})^{2}+\gamma ^{2}} 
\]
parametrized by four independently fluctuating parameters: the energetic
position $\epsilon _{0}$ of the resonance which is closest to the energy $%
\epsilon $, the location of the center of mass of the resonant state, $x$
and two inverse localization radia, $\alpha _{1}$ and $\alpha _{2}$. The
associated Wigner delay time can be represented as 
\[
\tau _{\phi }={\frac{1}{\gamma }}\frac{1}{1+\left[ (\epsilon -\epsilon
_{0})/\gamma \right] ^{2}}. 
\]

To analyze the distribution function $P(\tau _{\phi })$, we assume that the
center of the localized state and its energy have a homogeneous
distribution, and that the probability to find some value of the inverse
localization radius $\alpha _{i}$ in a segment of a wire with the length $%
x_{i}$ ( $x_{1}=x$ and $x_{2}=L-x$) is equal to 
\begin{equation}
P(\alpha _{i})=\sqrt{\frac{x_{i}}{2\pi \alpha _{0}}}\exp \left\{ -\frac{x_{i}%
}{2\alpha _{0}}(\alpha _{i}-\alpha _{0})^{2}\right\} .  \label{palphai}
\end{equation}
It is convenient to use random variables $p=\left( \alpha _{1}x_{1}+\alpha
_{2}x_{2}\right) /z$ and $q=\left( \alpha _{1}x_{1}-\alpha _{2}x_{2}\right)
/z$, instead of $\alpha _{1}$ and $\alpha _{2}$, where $z=L\alpha _{0}=L/4l$%
, which can be described using the joint distribution function 
\begin{equation}
P(p,q)=\frac{1}{2\pi }\frac{z}{\sqrt{1-y^{2}}}\exp \left\{ -\frac{z}{2}\left[
(p-1)^{2}-\frac{(q-yp)^{2}}{(1-y^{2})}\right] \right\} ,  \label{p-pq}
\end{equation}
where $y=(x/L)-2$ is uniformly distributed within the interval $[-1,1]$. In
the same parametrization, one can represent the Wigner delay time as 
\[
\tau _{\phi }=\frac{\tau }{1+\left( \epsilon /\gamma \right) ^{2}}\frac{\exp
(zp)}{\cosh (zq)},
\]
and the corresponding probability density as the conditional probability
integral 
\begin{eqnarray}
P(\tau _{\phi }) &=&\tau ^{-1}\int_{-1}^{1}dy\int dpdqP(p,q)
\label{defp-phi} \\
&&\times \int \frac{d\epsilon }{\Delta }\delta \left( \frac{1}{1+\left(
\epsilon /\gamma \right) ^{2}}\frac{\exp (zp)}{\cosh (zq)}-\frac{\tau _{\phi
}}{\tau }\right) .  \nonumber
\end{eqnarray}

To analyze Eq. (\ref{defp-phi}), we evaluate the integral using the
saddle-point method, and for $\tau _{\phi }<\tau e^{z}$ find that 
\begin{equation}
P(\tau _{\phi })\sim \frac{\tau }{\tau _{\phi }^{2}}.  \label{p-tauphi1}
\end{equation}
Note that the length of the chain does not appear in this result, which
indicates that this intermediate asymptotic of the delay-time distribution,
which is related to the distribution of the spectral width of the resonant
states, is dominated by the electron escape-rate from the resonant state
into the nearest reservoir and for $L\rightarrow \infty $ is exact for any
delay time. However, the finite length $L$ determines a cut-off, $\tau
_{\phi }\sim \tau e^{z}$ for this universal behavior, and for $\tau _{\phi
}>\tau \exp (z)$ we find 
\[
P(\tau _{\phi })\sim {\frac{\exp (-z/2)}{\tau }}\left( \frac{\tau }{\tau
_{\phi }}\right) ^{1+(1/(2z))\ln {(\tau _{\phi }/\tau )}}. 
\]
in agreement with Melnikov \cite{Melnikov81}.

To illustrate the validity of the above estimates, we compute the scattering
matrix of a series of equal strength, randomly spaced delta-function
scatterers. The spacings between scatterers possesses a Poisson
distribution, and the system is described by the Hamiltonian

\begin{equation}
-\frac{\hbar ^{2}}{2m}\frac{d^{2}}{dx^{2}}+U_{0}\sum_{i=1}^{L}\delta
(x-x_{i}).  \label{11}
\end{equation}

In the region $x_{j-1}<x<x_{j}$, an eigenstate of energy $\epsilon $ is of
the form 
\[
\psi _{\epsilon }(x)=A_{j}e^{ikx}+B_{j}e^{-ikx},\quad x_{j-1}<x<x_{j} 
\]
where $k=\sqrt{2m\epsilon/\hbar^2}$ and the wave amplitudes on either side
of the scatterer $j$ satisfy

\begin{equation}
\left( 
\begin{array}{c}
A_{j+1} \\ 
B_{j+1}
\end{array}
\right) =T_{j} \left( 
\begin{array}{c}
A_{j} \\ 
B_{j}
\end{array}
\right)
\end{equation}

In this expression $T_{j}$ is the transfer matrix,

\begin{equation}
T_{j}=\left( 
\begin{array}{cc}
1-i\beta & -i\beta e^{-2ikx_{j}} \\ 
i\beta e^{2ikx_{j}} & 1+i\beta
\end{array}
\right) .
\end{equation}
where $\beta =m\mu /{\hbar }^{2}k$.

The transfer matrix for a series of $N$ scatterers has the form,

\begin{equation}
{\tilde{T}}=\left( 
\begin{array}{cc}
{\tilde{T}_{11}} & {\tilde{T}_{12}} \\ 
{\tilde{T}_{21}} & {\tilde{T}_{22}}
\end{array}
\right) .
\end{equation}
and is given by the product of the transfer matrices of each individual
scatterer 
\[
{\tilde{T}}=T_{N}T_{N-1}\cdots T_{1}, 
\]
from which one obtains the scattering matrix

\begin{equation}
{\bf S=\left( 
\begin{array}{cc}
r & {t^{\prime }} \\ 
{t} & {r^{\prime }}
\end{array}
\right) .}
\end{equation}
via the relation

\begin{equation}
{\bf S=\left( 
\begin{array}{cc}
{-T_{21}T_{22}}^{-1} & {T_{22}}^{-1} \\ 
{T_{11}^{\ast }}^{-1} & {T_{12}T_{22}}^{-1}
\end{array}
\right) .}
\end{equation}
For a chain of length $L$ the conductance \ and the inverse localization
radius are determined by

\[
g=|t|^{2}\;{\rm and}\;\alpha {(L)}\sim \frac{ln|t|^{2}}{L}. 
\]

As an example of Eq. (\ref{p-alpha}), and to emphasize that we are
simulating a chain of weak scatterers for which $\alpha_0^{-1}$ is greater
than the mean spacing, Fig. 1 shows $P(\alpha )$ for a chain of scatterers
of concentration unity, strength $U_{0}=3$ and energy $\epsilon =4.1\pi ^{2}$%
. These same parameter values were used in all numerical simulations
described below.

Fig. 2 shows the corresponding plot of $\ln P(\tau _{\phi })\tau _{\phi
}^{2} $ for the chain lengths $L=50,70,100,200$. These results were obtained
by evaluating a finite difference $\tau _{\phi }=\left( \theta (\epsilon
+\delta )-\theta (\epsilon )\right) /\delta $, and were found to be stable
with respect to any choice of $\delta $ within the range $10^{-9}<\delta
<10^{-6}$. For $ln\tau >0$ all curves exhibit a plateau with a slope that
tends to zero with increasing length, in a good agreement with the
analytically estimated asymptotic behavior in Eq. (\ref{p-tauWigner}).

Having obtained agreement with known results for $P(\alpha )$ and $P(\tau
_{\phi })$, we now present an analysis of the transport time distribution, $%
P(\tau _{t})$, which represents the central new result of this paper. Using
the resonant tunneling description of Section II, the transport time for a
particle with a given energy $\epsilon $ is given by

\begin{equation}
\tau _{t}=T\tau _{\phi }=\frac{4\gamma _{1}\gamma _{2}\gamma ^{-3}}{\left( 1+%
\left[ (\epsilon -\epsilon _{0})/\gamma \right] ^{2}\right) ^{2}}.
\end{equation}
This allows us to parametrize the transport time using the position $x$ of
the resonant state center and two (left and right) inverse localization
radia, $\alpha _{1,2}$ as in Section II, 
\[
\tau _{t}=\frac{4\tau }{\left[ 1+\left( \epsilon /\gamma \right) ^{2}\right]
^{2}}\frac{\exp (zp)}{\cosh ^{3}(zq)}. 
\]
The probability to find a given value of $\tau _{t}$ can again be expressed
as a conditional probability 
\begin{eqnarray}
P(\tau _{t}) &=&\tau ^{-1}\int_{-1}^{1}dy\int dpdqP(p,q)  \label{ptransp1} \\
&&\times \int \frac{d\epsilon }{\Delta }\delta \left( \frac{4}{\left[
1+\left( \epsilon /\gamma \right) ^{2}\right] ^{2}}\frac{\exp (zp)}{\cosh
^{3}(zq)}-\frac{\tau _{t}}{\tau }\right) .  \nonumber
\end{eqnarray}
and evaluation of the integral in Eq. (\ref{ptransp1}) using the
saddle-point method yields the result of Eqs.(\ref{p-tauT1}-\ref{p-tauT3}).
Note that the power-law asymptotic $P(\tau _{t})\sim \tau
^{-1}e^{-z/2}\left( \frac{\tau }{\tau _{t}}\right) ^{4/3}$, valid for the
finite length wires within the parametric interval $-2/3<(1/z)\ln (\tau
_{t}/\tau )<1/3$, formally transforms into the universal central body of the
distribution in the thermodynamic limit $L\rightarrow \infty $.

To illustrate the validity of this result, Fig. 3 shows plots of the
function $\ln P(\tau _{t})\tau _{t}^{4/3}$ versus $\ln \tau _{t}$ for
various lengths. These numerical simulations show that at large $\tau _{t}$
all curves exhibit a plateau with a slope that tends to zero with increasing
length, demonstrating that the tail in the distribution of $\tau _{t}$,
varies as $\tau _{t}^{-4/3}$.

In summary, we have shown how earlier results for the Wigner-delay time $%
\tau _{\phi }$, based on a picture of resonant transport through localized
states, can be extended to yield the distribution of the transport time $%
\tau _{t}=T\tau _{\phi }$. In contrast with the distribution of $\tau _{\phi
}$, which exhibits a universal $1/\tau _{\phi }^{2}$ tail, the corresponding
intermediate asymptotic of the distribution of $\tau _{t}$ exhibits a
universal $1/\tau _{t}^{4/3}$ behavior.

\begin{figure}
{\psfig{figure=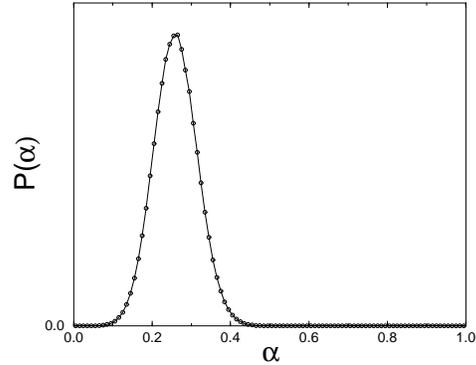,width=7cm}}
\caption{ Distribution $P(\protect\alpha )$ versus $\protect\alpha $ for a
chain length of 200 }
\label{f1}
\end{figure}

\begin{figure}
{\psfig{figure=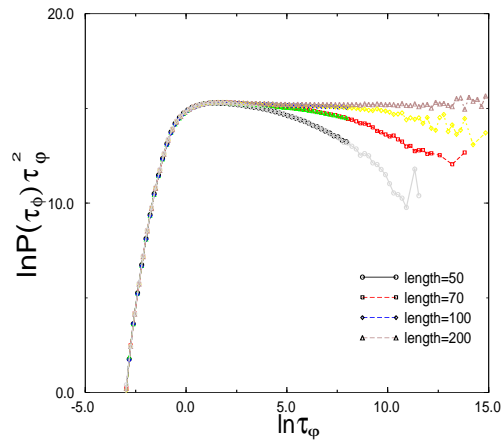,width=7cm,height=6.25cm}}
\caption{Plots of $lnP(\protect\tau_{\protect\phi} )\protect\tau_{\protect%
\phi}^{2}$ versus $ln\protect\tau $, for various lengths, , ranging from $L=50$ (upper curve) to $L=200$ (lower curve. The size of the
ensemble is $10^{8}$}
\label{f2}
\end{figure}

\begin{figure}[tbp]
{\psfig{figure=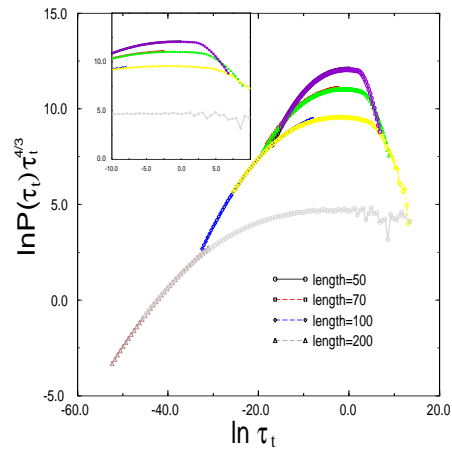,width=7cm,height=6.25cm}}
\caption{Plots of $\ln P(\protect\tau_t)\protect\tau_t^{4/3}$ versus $\ln%
\protect\tau_t$, for various lengths, ranging from $L=50$ (upper curve) to 
$L=200$ (lower curve. The insert shows the central portions of the
distributions. The size of the ensemble is $10^{8}$}
\label{f4}
\end{figure}


\begin{references}
\bibitem{Altshuler96}  B.L. Altshuler and B.D. Simons, in Mesoscopic Quantum
Physics, edited by E. Akkermans, G. Montabaux, J.-L. Pichard, and J.
Zinn-Justin (North-Holland, Amsterdam, 1996).

\bibitem{Mehta91}  M. L. Mehta, {\it Random Matrices and Statistical Theory
of Energy Levels} (Academic, New York, 1991).

\bibitem{Jalabert92}  R. A. Jalabert, A. D. Stone, and Y. Alhassid, Phys.
Rev. Lett. {\bf 68}, 3468 (1992).

\bibitem{Fyodorov96}  Ya.V. Fyodorov and H-J. Sommers, Phys. Rev. Lett, {\bf %
76} 4709 (1996)

\bibitem{Fisher80}  P. W. Anderson, D. J. Thouless, E. Abrahams, and D. S.
Fisher, Phys. Rev. B {\bf 22}, 3519 (1980).

\bibitem{Melnikov81}  V.I. Melnikov, Sov. Phys. Solid State {\bf 23}, 444
(1981) [Fiz. Tverd. Tela {\bf 23}, 782 (1981)].

\bibitem{Landauer94}  R. Landauer and Th. Martin, Rev. Mod. Phys. {\bf 66},
217 (1994).

\bibitem{Gasparian95}  V. Gasparian and M. Pollak, Phys. Rev. {\bf B} {\bf 47%
}, 2038 (1993).

\bibitem{WignTimeDisrt}  C. Texier and A. Comtet, cond-mat 9812196, Dec
(1998)

\bibitem{BeenakkerReview}  C.W.J. Beenakker, Rev. Mod. Phys. {\bf 69} 731
(1997)

\bibitem{Pastawski92}  H. M. Pastawski, Phys. Rev. {\bf B} {\bf 46}, 4053
(1992).

\bibitem{Buttiker93}  M. B\"{u}ttiker, A. Pr\^{e}tre, and H. Thomas, Phys.
Rev. Lett. {\bf 70}, 4114 (1993).

\bibitem{Buttiker96}  V.A. Gospar, P.A. Mello and M. B\"{u}ttiker Rev. Lett. 
{\bf 77}, 3005 (1996).

\bibitem{Kohlman96}  R.S. Kohlman, J. Joo, Y.G. Min, A.G. MacDiarmid, and
A.J. Epstein, Phys. Rev. Lett. {\bf 77}, 2766 (1996).

\end{references}
\end{document}